%% file: main.tex
\documentclass[]{interact}

\usepackage{amsmath}
\usepackage{amssymb}
\usepackage{graphicx,psfrag,epsf}
\usepackage{enumerate}
\usepackage{url} 
\usepackage[dvipsnames]{xcolor}
\usepackage{tikz}
\usepackage{xparse}
\usepackage{multirow}
\usepackage{float}

\usepackage{epstopdf}
\usepackage[caption=false]{subfig}

\usepackage[numbers,sort&compress]{natbib}
\bibpunct[, ]{[}{]}{,}{n}{,}{,}
\renewcommand\bibfont{\fontsize{10}{12}\selectfont}

\definecolor{darkgreen}{rgb}{0.0, 0.5, 0.0}

\newcommand{\blind}{1}
\newcommand{\supplement}{0}

\theoremstyle{plain}

\theoremstyle{definition}

\theoremstyle{remark}

\begin{document}

\setcitestyle{square}


\if0\supplement
{
\if1\blind
{
\date{}
  \title{\bf 
    Partially Constrained Group Variable Selection to Adjust for Complementary Unit Performance in American College Football
  }
  \author{\name{A. Skripnikov\textsuperscript{a}\thanks{CONTACT A. Skripnikov. Email: askripnikov@ncf.edu}}
  \affil{\textsuperscript{a}New College of Florida, Sarasota, FL 34243, USA}}
  
 \maketitle
 
} \fi

\if0\blind
{

\date{}
  \title{Partially Constrained Group Variable Selection to Adjust for Complementary Unit Performance in American College Football
  }
  
  \maketitle
  
} \fi

} \fi

\if1\supplement
{
\date{}

  \title{Supplementary Materials for "Partially Constrained Group Variable Selection to Adjust for Complementary Unit Performance in American College Football"
  }
  
   \maketitle
   
} \fi

\if0\supplement
{

\if0\blind{

\begin{abstract}
\input{Arxiv_Submission/1_Abstract.tex}
\end{abstract}

} \fi

\if1\blind{

\begin{abstract}
\input{Arxiv_Submission/1_Abstract_Unblinded.tex}
\end{abstract}

} \fi

\begin{keywords}
group penalty, LASSO, natural splines, regularized estimation, reverse causality, sports statistics
\end{keywords}

\input{Arxiv_Submission/2_Introduction.tex}
\input{Arxiv_Submission/3_Materials_and_Methods}

\input{Arxiv_Submission/4_Results}

\input{Arxiv_Submission/5_Discussion}

\input{Arxiv_Submission/6_Acknowledgments}

\input{Arxiv_Submission/7_Declaration_of_Interest}
\input{Arxiv_Submission/8_Appendices}

\bibliographystyle{agsm}
\bibliography{bibliography}

} \fi

\end{document}

%% file: Arxiv_Submission/1_Abstract_Unblinded.tex
Given the importance of accurate team rankings in American college football (CFB) - due to heavy title and playoff implications - strides have been made to improve evaluation metrics across statistical categories, going from basic averages (e.g. points scored per game) to metrics that adjust for a team's strength of schedule, but one aspect that hasn't been emphasized is the complementary nature of American football. Despite the same team's offensive and defensive units typically consisting of separate player sets, some aspects of your team's defensive (offensive) performance may affect the complementary side: turnovers forced by your defense could lead to easier scoring chances for your offense, while your offense's ability to control the clock may help your defense. For 2009-2019 CFB seasons\footnote{Data and source code files are made available at \url{https://github.com/UsDAnDreS/SUBMISSIONS_OffenseST_DefenseST_GLM_LASSO_adjusted_rankings}}, we incorporate natural splines with group penalty approaches to identify the most consistently influential features of complementary football in a data-driven way, conducting partially constrained optimization in order to additionally guarantee the full adjustment for strength of schedule and homefield factor. We touch on the issues arising due to reverse-causal nature of certain within-game dynamics, discussing several potential remedies. Lastly, game outcome prediction performances are compared across several ranking adjustment approaches for method validation purposes.

%% file: Arxiv_Submission/2_Introduction.tex
\section{Introduction}
\label{sec:Introduction}

American football, while requiring immense physical ability, is an example of an extremely strategic sport. Requiring full mental engagement on every play from all 11 players your team puts on the field, it underlines the importance of tactical preparation and accurate evaluation of the opposing team in advance of the match-up. Besides that, in American college football specifically, the ability to objectively evaluate team performance is pivotal for the purposes of rankings, with the latter fully dictating the college teams that get into the championship playoffs and high-profile bowl games, including all the financial benefits to come with it. Historically, classical averages have been used to evaluate team's performance in a certain statistical category \citep{albert2005anthology}. For example, American football team's offensive output from yardage standpoint is traditionally measured in total yards the team gains per game; its tendency to turn the ball over to the opposing team is typically captured in their average turnover count per contest; team's pure scoring ability is ubiquitously described by an average amount of points scored per game. While useful, such basic averages are extremely flawed whenever trying to objectively evaluate and compare teams within the setup of Football Bowl Subdivision (FBS) of American college football. 

Football Bowl Subdivision (FBS), as of year 2019, consists of $131$ participating college football programs, split into $11$ conferences. Each team is typically scheduled to play only $12$ opponents during the regular season, with most of the games taking place against the teams from the same conference, hence leaving the ranking picture incomplete in terms of relative strengths for teams across the entire FBS. Certain conferences could end up being especially strong, or weak, during a particular season, leading to uneven quality of opposition which can't be reflected in the calculating each team's basic per-game averages alone. That gave rise to what's now known as "strength of schedule" adjustment, originating in \citep{harville1977use} via an offense-defense model where points scored by a team are assumed to be a function that team's offensive strength and opposing team's defensive strength, therefore adjusting for quality of the opponent. 
Lastly, given the importance of home-field advantage \citep{edwards1979home, vergin1999no}, in addition to adjusting for strength of scheduled opponents \citep{harville1977use} also accounted for whether a team had a home- or road-heavy schedule.

In this work, besides the strength of schedule and homefield adjustments, we wanted to also incorporate the complementary nature of American football. Unlike most other sports, in American football offense and defense are played by separate, in most cases non-overlapping, units of players, that can't be on the field at the same time. It might lead to an assumption that performance of a team's offense could be treated independently of that same team's defense. Nonetheless, historically it's been shown that the two sides are likely to complement each other, e.g. turnovers forced by your defense could lead to easier scoring chances for your offense, while your offense's ability to control the clock may help your defense. That concept has been termed "complementary football", and focus of this work was on determining the most critical features of collegiate American football in affecting the complementary side's performance. Specifically, we utilized a variety of regularized estimation techniques that impose a penalty for the purposes of variable selection, implementing partially constrained optimization with the goal of guaranteeing strength of schedule and homefield adjustment. Somewhat related ideas were studied in \citep{karlis2003analysis, boshnakov2017bivariate} but in application to modelling the dependence between goals scored and allowed by a team in a soccer match. To the best of our knowledge, there hasn't been any involved research done on complementary impacts an American football offense can have on its own defense, and vice versa, after having adjusted for strength of schedule and homefield factor.

The remainder of the paper is structured as follows. Section \ref{sec:MaterialsMethods} lays out the details of data collection and cleaning, formulates all regularized estimation approaches for the purpose of finding the most critical complementary football features in affecting certain statistical categories. The results of feature selection, ensuing ranking adjustments, along with their impact on predicting game outcomes, are presented in Section \ref{sec:Results}. Moreover, the issue of endogeneity is brought up, resulting from reverse causal within-game dynamics of a football game, and several partial solutions discussed, main one involving the use of solely efficiency-based statistics. Section \ref{sec:Discussion} contains concluding remarks and discussions of future work.

%% file: Arxiv_Submission/3_Materials_and_Methods.tex
\section{Materials and methods}
\label{sec:MaterialsMethods}

\subsection{Data collection and cleaning}
\label{sec:DataCollectionCleaning}

Game-by-game data for 2009-2019 FBS seasons was obtained via web scraping from two primary sources: \url{www.sports-reference.com/cfb/} and \url{www.cfbstats.com/}. The former website provides detailed scoring and defensive play, and has webpage format conducive to seamless scraping of analysis-ready data tables into R Statistical Software \citep{RCoreTeam} - primary tool used for statistical analysis in this work - via $rvest$ package \citep{rvest}. 
The latter website helped augment the data with more statistical measurements, 
although it required a more profound web scraping effort while using 
$RSelenium$ package \citep{RSelenium}.
Certain defensive play statistics (e.g. tackles, sacks, forced fumbles) were lacking for "non-major" teams, meaning colleges not in the FBS. 
Despite rare occurrence of such observations (only about $7\%$), we wanted to incorporate all relevant games into our analysis 
and, instead of removing them, we performed data imputation \citep{efron1994missing} via linear regression. We regressed the missing statistics - e.g. sacks - on the other statistical categories that were available for non-major teams (e.g. yards, touchdowns, punts, etc), utilizing observations for all FBS teams in fitting this regression, to later make predictions of that missing statistic for non-major teams.

Having obtained game-by-game data on teams' performances in a multitude of statistical categories, we converted it into a format where each observation describes the cumulative performance throughout the game whenever one team's offensive side (offense or offensive special teams) - and, consequently, opponent's defensive side (defense or defensive special teams) - was on the field. E.g. for this team's offense, it would include yards gained, touchdowns scored, passes completed, field goals scored, etc, while for opponent's defense it would show total tackles, tackles for loss, sacks, quarterback hurries, etc. To clarify, "offensive special teams" imply the field goal kicking and kick/punt returning player sets, while "defensive special teams" represent players participating in field goal blocking and kick/punt coverage. 
That data format results into two observations per each game, one representing plays produced when first team's offensive side (and second team's defensive side) was on the field, second - when second team's offensive side (and first team's defensive side) was on the field. This implies potential within-game dependence introduced by such pairs of observations, but after having calculated Intraclass Correlation Coefficients \citep{bartko1966intraclass} across the models considered in this paper, virtually none of those correlations ended up being over 0.10 in magnitude, pointing to independence as a reasonable assumption for our data.

\subsection{Variable selection methodology outline}
\label{sec:VariableSelectionOutline}

Below we introduce the modeling notation and main variable selection approaches to help adjust team's offensive (or defensive) quality for the strength of schedule, homefield factor, and complementary nature of American football.



\subsubsection{Notation}
\label{sec:Notation}

Assume we have a total of $n$ teams in the league, and we are interested in a particular statistical category $y$, be it yards (per game), turnovers, points, etc. Let $y_{ij}$ denote the performance of team $i$ against team $j$ in category $y$, $n_i$ - total number of opponents team $i$ has had through the season, $i,j =1,\dots,n$. For example, if $y=$ {\em \{total   yards gained per  game\}}, then $y_{ij}$ would correspond to the amount of yards team $i$ gained when playing against team $j$, and, by symmetry, the amount of yards team $j$ allowed in the game against team $i$. In case teams $i$ and $j$ had to face each other $L_{ij}$ times during the season, we add another index $l$. Then, $y_{ijl}$ corresponds to performance of team $i$ in category $x$ during its $l^{th}$ meeting with team $j$, $i,j=1,\dots,n, \ l=1,\dots,L_{ij}$.

Next, let $h_{ijl}$ denote a homefield indicator for $l^{th}$ game between teams $i$ and $j$, taking on value $1$ if team $i$ is at home, $0$ if the game site is neutral, $-1$ if team $i$ is at home. Such numerical encoding was intuitive (typically easier to play at home than at a neutral site, and at a neutral site than on the road) and also got confirmed by running a dummy-variable encoding scheme, having shown increases (in points, touchdowns scored, yards gained) for home and decreases for road games compared to the neutral site baseline. 


To incorporate adjustment for strength of the schedule, we introduce concepts of offensive (defensive) worth of the "league-average opponent", and offensive (defensive) margin for a team. In regards to a particular statistical category $y$, one can define the league-average opponent via two parameters -  offensive and defensive worth. 
E.g. for points per game, offensive (defensive) worth of the average opponent is the average points per game scored (allowed) by all teams across all games that could have been played against one another throughout the course of the season. Due to symmetry (team $i$ scoring $y_{ij}$ points against team $j$ is equivalent to team $j$ allowing $y_{ij}$ points to team $i$), both offensive and defensive worth represent the same value for the league-average opponent, which we denote as $\mu$. 
Now, for each team $i$ we can posit parameters capturing two aspects of its performance within a statistical category -  offensive margin $\alpha_i$ and defensive margin $\beta_i$. Offensive (defensive) margin describes by how much a team would outperform the aforementioned defensive (offensive) worth $\mu$ of the average opponent. The main assumption when adjusting for strength of schedule is that performance of team $i$ against team $j$ in category $y$ is attributable to both the offensive margin $\alpha_i$ of team $i$ and defensive margin $\beta_j$ of team $j$.

Lastly, presuming that we consider $C$ complementary football statistics, let's use $x_{c,jil}, \ c=1,\dots,C,$ to denote the value of $c^{th}$ statistic that's complementary to the $y_{ijl}$, meaning that $x_{c,jil}$ is obtained when the defense (complementary unit for the offense) of $i^{th}$ team and offense (complementary unit for the defense) of $j^{th}$ team were on the field during their $l^{th}$ game of the season between these two teams. 

\subsubsection{Natural cubic splines}
To model potentially non-linear effects of complementary football features, as a well-known method we utilized natural cubic splines \citep{hastie_09_elements-of.statistical-learning}, where one uses a mixture of piece-wise cubic and linear polynomials, smoothly connected at a set of $K$ knots placed across the range of the explanatory variable. It results into each complementary statistic $x_c$ being represented by a set of basis functions $N_1(x_c), N_2(x_c),\dots, N_{K-1}(x_c)$.
For more detail, see \citep{hastie_09_elements-of.statistical-learning}, keeping in mind that the intercept basis function $N(x_c)=1$ for each individual complementary statistic $x_c$ in our case is omitted from the basis due to being folded into the overall model's intercept. We chose to use $K=5$ knots placed at $0.00$-, $0.25$-, $0.50$-, $0.75$- and $1.00$-quantiles, providing just enough flexibility to capture any clear non-linearity, while decreasing chances of overfitting and low interpretability that come with overly flexible fits. That results into each complementary statistic $x_c$ being represented by four basis functions, with its partial effect on response calculated via a linear combination of these functions.


\subsubsection{Partially constrained group penalization}

Natural splines approach leads to each complementary football feature being represented by a group of several parameters, where inclusion of a feature into the model implies including the entire said parameter group, while exclusion indicates setting that whole group of parameters to zero. Such setting warrants use of group penalization approaches \citep{yuan2006model, huang2012selective, wang2008note}. 
As we also want to guarantee adjustment for homefield effect and strength of schedule,  certain parameters will be left unpenalized, resulting into the following general format of partially constrained group penalization criteria: 

\begin{multline}
    \min_{\substack{\mu, \delta, \{\alpha_i\}_1^n, \{\beta_j\}_1^n  \\ \{\pmb{\gamma}_{c}\}_1^{K-1}, \ c=1,\dots,C}} \ \sum_{i, j=1}^n \sum_{l=1}^{L_{ij}} (y_{ijl} - (\mu + \delta h_{ijl} + \alpha_i + \beta_j  + \sum_{k=1}^{K-1} \gamma_{1,k} N_k(x_{1,jil}) + \dots + \sum_{k=1}^{K-1} \gamma_{C,k} N_k(x_{C,jil})))^2  + \\ \lambda \sum_{c=1}^C P_c( ||\pmb{\gamma}_c||_2),
    \label{eq:GroupPenaltyCriter}
\end{multline}
where $\pmb{\gamma}_c = (\gamma_{c,1}, \dots, \gamma_{c,K-1})$, capturing all the natural cubic spline coefficients for $c^{th}$ complementary statistic; $||\pmb{\gamma}_c||_2 = \sqrt{\sum_{k=1}^{K-1} \gamma_{c,k}^2}$ - the Eucledian ($L_2$) norm; $P_c(\cdot)$ indicates a specific group penalty function to be utilized for the coefficients of $c^{th}$ complementary statistic; $\lambda$ - tuning parameter responsible for strength of penalty (larger $\lambda$ implies larger penalty for high values of $||\pmb{\gamma}_c||_2$). Here, the penalty is imposed solely on coefficients $\pmb{\gamma}_1, \dots, \pmb{\gamma}_C$ for complementary football statistics, with the goal of inferring the most critical ones for statistical adjustment purposes. This partially constrained optimization criteria guarantees inclusion of homefield effect ($\delta$) and strength of schedule adjustment effects ($\alpha_i, \beta_j, \ i,j=1,\dots,n$).

We considered an array of various group penalties, to subsequently identify the most consistently selected features regardless of a particular penalty choice. First, we utilized a classic group least absolute shrinkage and selection operator (group LASSO, \cite{yuan2006model}), $P_c(||\pmb{\gamma}_c||_2) = ||\pmb{\gamma}_c||_2$, which is a natural extension of regular LASSO \citep{tibshirani1996regression} and selects variables in a grouped manner. Due to applying the same strength of penalty ($\lambda$) to all coefficients, regular LASSO was shown to struggle with efficiency \citep{fan2001variable} and model selection  \citep{leng2006note}. Given the shared nature of the penalty function, similar issues persist for group LASSO. In case of regular LASSO, adaptive LASSO \citep{zou2006adaptive} method was proposed to address this issue via introducing coefficient-specific weights, subsequently yielding consistent variable selection and estimators satisfying oracle property. 
The exact same intuition was carried out in \citep{wang2008note} to implement adaptive group LASSO. In our case, it leads to penalty function $P_c(||\pmb{\gamma}_c||_2) = w_c ||\pmb{\gamma}_c||_2, \ w_c = 1/||\hat{\pmb{\gamma}}^{LS}_c||^{\gamma}_2$, where $\hat{\pmb{\gamma}}^{LS}_c$ denotes the least squares estimate for the effect of $c^{th}$ complementary statistic resulting from optimizing criteria (\ref{eq:GroupPenaltyCriter}), but with the penalty term ($\lambda \sum_{c=1}^C P_c( ||\pmb{\gamma}_c||_2)$)  excluded. 
While adaptive LASSO poses a two-stage approach, with estimation of weights $\{w_c\}$ preceding the actual penalization, non-convex penalties present a single-stage alternative that achieves analogous bias reduction for large regression coefficients. We used two of the most popular non-convex penalties - smoothly clipped absolute deviations (SCAD) \citep{fan2001variable} and minimax concave penalty (MCP) \citep{zhang2010nearly} - which were originally defined for regular LASSO, but can be naturally extended to group LASSO setting. For details on the exact functional form $P_c(\cdot)$ for those penalties see \citep{huang2012selective}. Three aforementioned penalties - adaptive, SCAD, MCP - have an extra tuning parameter (generally denoted as  "$\gamma$" in the original papers), that in each case we picked in a data-driven way according to Bayesian Information Criterion (BIC) \citep{neath2012bayesian}. 

To implement all of the aforementioned group LASSO penalties for optimization task (\ref{eq:GroupPenaltyCriter}) we utilized $grpreg$ package \cite{breheny2021package}, along with 20-fold cross-validation (CV) \citep{browne2000cross} and BIC \citep{neath2012bayesian} to select several candidate values for the tuning parameter $\lambda$. For CV, we opted to use a $\lambda$ value that achieved CV test error within one standard deviation of the minimal CV test error, while providing a sparser model (selecting less variables). Moreover, to reduce the randomness effect of the 20-fold CV procedure, we ran it five times for each season, only counting features that got selected at least four out of five times. We used BIC and sparse CV $\lambda$-value 
because, amongst all of the most ubiquitous criteria, these ones imposed strongest penalties on overly complicated models, leading to a small set of the most important features.  Most importantly, if a complementary football feature gets selected by these criteria, it is virtually guaranteed to also be selected by any other ubiquitous criteria.

Each complementary statistic was evaluated on consistency of its selection across 2009-2019 FBS seasons by each of four main variable selection methodologies outlined in Section \ref{sec:MaterialsMethods} (classic and adaptive group LASSO, group SCAD and MCP penalties). Specifically, each method was given equal weight of $\frac{1}{4}$ in the final consistency metric, while within any individual method an equal weight of $\frac{1}{2}$ was granted to each of the two considered tuning parameter selection approaches (BIC and sparse CV $\lambda$-value).

Lastly, to calculate the final adjustment for each season, natural cubic splines were fitted solely with the complementary football features that were consistently picked across variable selection methods, while centering variables representing complementary statistics around their respective means. That way, team $i$'s offensive worth $\mu + \alpha_i$ and defensive worth $\mu + \beta_i$ will be projected not only onto league-average opponent and neutral site, but also onto league-average complementary side.

\subsubsection{Modeling assumptions given response variable type}
\label{sec:ResponseModelingAssumptions}

We considered six statistics to be adjusted, breaking it down by the type of a statistic (points, touchdowns, yards) and whether "counter-plays" are being accounted for or not. "Counter-plays" are plays that happen  directly after a turnover, before (if at all) the complementary side takes the field. For example, we could solely count touchdowns that your offense scored, or we could also take the margin between touchdowns they scored and touchdowns they allowed the opposing defense to score (via a returned interception or fumble). Same for points and yards.
The primary reason for creating these margin-statistics is to eventually use them as features to forecast game outcomes, given that they account for totality of the impact whenever a particular side took the field, rather than reflecting solely their positive impact. 
Lastly, due to the discrete, low-count, nature of the touchdowns category (while points and yards are well-approximated by continuous distributions), we also considered tailored count response approaches (e.g. Poisson and negative binomial generalized linear models), but classic linear regression yielded a fit of either superior or similar quality to that of the alternatives, hence was picked moving forward.


%% file: Arxiv_Submission/4_Results.tex
\section{Results}
\label{sec:Results}

This section is broken up into two main parts based on the complementary football feature sets being considered. First part works with the entire set of statistical categories originally obtained from \url{www.sports-reference.com/cfb/} and \url{www.cfbstats.com/}. Second part discusses using an efficiency-based feature set of solely per-play and per-possession statistics as a way to partially address the issue of reverse causality permeating the first set. 
Both feature sets can be found in the Appendix.

\subsection{Using entire feature set of complementary football statistics.}
\label{sec:UsingEntireFeatureSet}

\subsubsection{Collinearity.}
\label{sec:Collinearity}

Prioritizing interpretation and consistency of variable selection results, we disposed of both perfect and near-perfect multicollinearity amongst our features. For perfect multi-collinearity cases of one statistical category representing a sum of sub-categories (e.g. passing attempts equals completions plus incompletions) we chose to go with the sub-categories (e.g. retain completions and incompletions, while dropping passing attempts). The only exceptions were the cases where breakdown into sub-categories was not useful in terms of its effects on the complementary side. For example, as long as it is a turnover, distinguishing between fumbles and interceptions is unnecessary, as the impact on the complementary side won't change based solely on that aspect. On the other hand, there is a considerable difference between scoring turnovers (when defense instantly scores) and non-scoring ones (complementary side gets the ball back in the latter case, doesn't in the former), or between pass completions and incompletions (the latter stops the game clock, the former - doesn't). For near-perfect multi-collinearity cases, predominantly represented by strong pairwise correlations (over 0.90 in magnitude), we picked the more ubiquitously used statistic to represent the highly correlated group. For example, points per game being are strongly correlated with touchdowns per game (consistently over 0.95 each season), and we pick points per game as a more all-encompassing and recognizable statistic. For the feature set resulting from this step, which ended up containing roughly $50$ variables, see the Appendix.

\subsubsection{Selected complementary statistics, issue of reverse causality.}
\label{sec:ReverseCausality}

Among statistical categories that were consistently selected as affecting team's complementary side in regards to points scored/allowed in a game, we got rushing attempts, pass incompletions,  non-scoring turnovers, number of attempted special teams returns of the ball, and yards gained on those returns. Some of these are perfectly reasonable, e.g. non-scoring turnovers provide the complementary side with the ball and a potentially easy scoring opportunity. For a statistic such as pass incompletions, while expectedly having a positive impact on the points scored on complementary side (stopping the clock, hence leaving more game time to complementary side), it rarely features in football discussions of the most pivotal factors driving complementary scoring. Lastly, several statistical categories simply don't make as much sense, e.g. number of attempted returns or rushes, regardless of yards gained on such plays.

Why do we observe such unintuitive strong effects for some of the statistics? Reverse causality is the most likely answer. For example, while pass incompletions facilitate complementary side scoring, they can also be a partial function of play selection by the team that's behind in the score. Such team would be forced to throw the ball more often in order to gain yards quicker and catch up, while also stopping the clock in case of an unsuccessful play (unlike an ineffective rushing attempt, an incompleted pass actually stops the clock), trying to score while taking as little of the game clock as possible. For rushing attempts, on the other hand, reverse logic applies in terms of play selection: the leading team is likely to call more rushing plays to keep the game clock running, while the trailing team is less likely to do so, creating a reverse-causal component in estimating the effect of that complementary football feature. Lastly, the special teams return play data (attempts, yards) suffers from the same issue, but has an even clearer time-sequential component to it: a return attempt tends to follow  directly after someone having just scored on the complementary side.
Despite lack of access to sequential play-by-play data, game-by-game data allowed to calculate that 68\% of attempted returns followed after a kickoff, which in its turn is most likely to come after a score on the complementary side (60\% chance it was a 7-point touchdown, 18\% chance - a 3-point field goal). This clearly shows a reverse-causal pattern of a special teams return following a complementary side score, not the other way around.

How can we potentially alleviate these reverse-causal effects? There are several potential solutions, the most obvious one being to obtain play-by-play data across all games. With careful consideration and proper domain understanding, knowledge of play sequencing could provide us with further insight into the causal flow of within-game dynamics. At the very least, it could make clear that return attempts by one's offense typically directly follow after the score by the opposing offense, avoiding the aforementioned reverse-causal trap. Unfortunately, in many cases one lacks the capability to access and process such fine-grained play-by-play data, and therefore is left looking for more feasible alternatives.

Second approach could be adjusting for endogeneity, which broadly refers to cases of an explanatory variable being correlated with an error term, hence violating one of the least squares modeling assumptions and subsequently biasing the estimates.
Reverse causality is a special case of endogeneity, with explanatory variables suffering from it being called "endogenous",  due to being determined by factors "within the system" (e.g. within the game situation), as opposed to "exogenous", which are fully defined by external factors (e.g. weather). Issue of endogeneity has been discussed across several application domains \citep{kmenta1986elements, duncan2004endogeneity, rutz2019endogeneity}, with instrumental variables \citep{bowden1990instrumental} proposed as one of the most ubiquitous remedial approaches. A strong instrumental is an exogenous variable that strongly correlates with the endogenous explanatory variable, while being uncorrelated with the error term. Such variables have been notoriously difficult to find, resulting into cases where weak instruments only exacerbated estimation bias  \citep{bound1995problems}.  Moreover, in our specific case, virtually every complementary football feature can be considered endogenous due to being a partial byproduct of within-game dynamics, making it that much harder to find good exogenous instrumental variables, which left us looking for other alternatives.

Lastly, with unavailability of play-by-play data and complexity of endogeneity adjustment logistics, we decided to pursue an intuition-driven approach of focusing solely on efficiency statistics. While such cumulative metrics as pass incompletions or rushing attempts can be a strong function of strategic play selection within the context of the game, efficiency-based statistics are more reflective of team's effectiveness once a certain play was called, rather than heavily depending on the overall game context. For example, if a team pursues a pass-heavy strategy in an effort to catch up with the opponent, their cumulative passing statistics - attempts, completions, incompletions, yards - will inevitably increase, while their efficiency-based alternatives - completion percentage, yards per attempt - are not guaranteed to change one way or the other. Hence, to reduce the chance of reverse causality, we opt for using solely efficiency-based statistical categories, showing how successful a team is once a certain play was called, as opposed to getting overly affected by contextual strategic play choices.

\subsection{Using only efficiency-based complementary football statistics}
\label{sec:UsingOnlyEfficiencyBasedFeatures}

\subsubsection{Efficiency-based features}
\label{sec:EfficiencyBasedFeatureDescription}

\begin{figure}[h]
\begin{center}
 \includegraphics[scale=0.8]{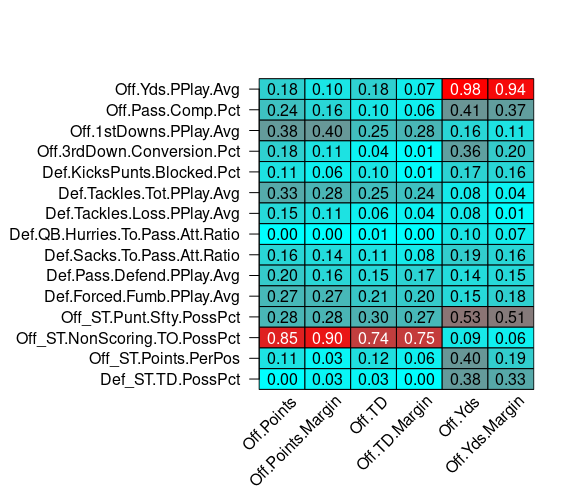}
\end{center}
\caption{\textbf{A weighted proportion of times a natural cubic spline representing a particular complementary football statistic (rows) was selected across different group penalties and tuning parameter selection approaches.}
}
\label{fig: EfficiencyConsistencySelection}
\end{figure}

Having disposed of collinearity issues and unnecessary statistical category breakdowns (see Section \ref{sec:Collinearity} for analogous reasoning), in the row names of Figure \ref{fig: EfficiencyConsistencySelection} one can witness the efficiency-based complementary football statistics that ended up being utilized for variable selection. For each statistical category we had to decide between using a per-play or per-possession efficiency. We selected to look at plays that terminate a possession (turnover, points scored, punt, etc) on per-possession basis, making them into possession percentages. For example, $Off\_ST.NonScoring.TO.PossPct$ corresponds to the percentage of possessions ending in the team's offensive side turning the ball over to their defensive side (as opposed to turning it over for an instant score by the opposing defensive side, which is captured via $Def\_ST.TD.PossPct$). For offensive scoring, we decided against breaking it down scoring possessions into touchdowns and field goals, in big part due to both outcomes resulting into same direct impact on the complementary side (a kickoff). Nonetheless, percentage of scoring (and non-scoring) possessions was heavily correlated with a more recognizable and informative metric of points per possession, hence was dropped in favor of the latter. Lastly, punts and safeties were combined into a single category because, despite safeties also resulting into two points scored (unlike zero points in case of punts), the biggest advantage of either play is the possession turning over to the other side, along with the similarity of the ball getting punted across the field in both cases.

On the other hand, virtually all other metrics were treated on per-play basis, including yards or first downs gained, overall tackles, tackles for loss, forced fumbles. For yards gained, there were several alternative considerations to simply looking at average yards per play by the offense: breaking it down into yards per pass and rush attempt, or also including special teams returns as plays into the overall per-play yardage. The former idea was tested via cross-validation, having shown that total yards per play outperformed the pass/rush yardage breakdown approach.  The latter idea resulted into special teams return yardage overly affecting the per-play efficiency values, leading to this statistic getting selected every single time, mostly due to the aforementioned reverse causality issue pertaining to return play data. Meanwhile, some other statistics would result into percentages due to their binary nature, e.g. pass completion, third down conversion, blocked kicks/punts.  One may notice that we didn't include statistics such as field goal percentage, average yards per punt or per attempted return, red zone conversion percentage, etc. That is due to the issue of a low (sometimes zero) denominator, where efficiency is calculated over a really small sample size within a game (e.g. only over two field goal attempts), and hence is not necessarily representative of team's true ability in that category, or the statistic's impact on the game. Moreover, in relatively frequent cases when sample size is zero, the value has to be set to zero or left undefined, neither option being optimal.

Lastly, such statistical categories as "sacks to pass attempts ratio" were defined in an effort to adjust for effects of pass-heavy play calling being conducive to such defensive plays as sacks or quarterback hurries. While a sack is not recorded as a passing attempt, and ideally one needs the number of quarterback dropbacks to create a proper efficiency metric, due to data limitations this was a reasonable alternative. In similar vein, percentage of forced fumbles by defense was only counted for rushing attempts and pass completions - the only plays where a fumble was possible.

\begin{figure}[H]
\begin{center}
 \includegraphics[scale=0.45]{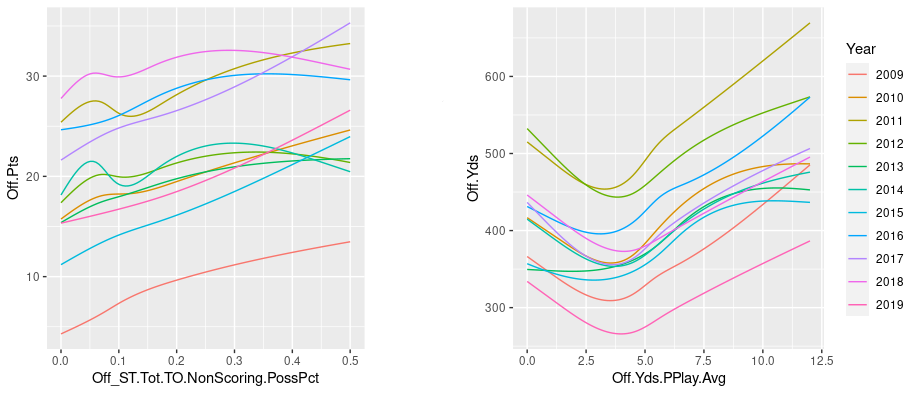}
\end{center}
\caption{\textbf{Non-linear effects of complementary statistics on respective response categories, controlling for strength of schedule and homefield factor. Left: Effect of non-scoring turnovers forced by your defense on points scored by your offense. Right: Effect of per-play yardage allowed by your defense on yards gained by your offense.} These particular effect graphics were obtained for a Navy versus Army match-up at a neutral site. Changing the match-up would only affect the intercepts, hence there's no loss of generality in depicting the nature of the effect.
}
\label{fig: NonLinearEffects_Graphs_Efficiency_Stats}
\end{figure}

\subsubsection{Selected complementary statistics.}
\label{sec:SelectedEfficiencyStatistics}

Figure \ref{fig: EfficiencyConsistencySelection} demonstrates the efficiency-based features of complementary football (rows) that affected each respective statistical category to be adjusted (columns). As described in Section \ref{sec:ResponseModelingAssumptions},  "margin" accounts for counter-plays in a respective category.

One can witness nonscoring turnover percentage forced by your defensive side (or committed by opponent's offensive side, hence the "$Off\_ST$" prefix) as the sole statistic that is consistently picked ($\geq 74\%$ of the time) in terms of impacting points and touchdowns scored by your (complementary) offensive side. Meanwhile, for both total yardage response categories, the only complementary football statistic picked with strong stability is the total yards per play ($\geq 94\%$ of the time). To evaluate the nature of these consistently selected complementary statistics, we fit the respective natural cubic spline regressions, with Figure \ref{fig: NonLinearEffects_Graphs_Efficiency_Stats} depicting the effects of your defense's forced non-scoring turnovers on your offensive scoring (left), and of per-play yardage allowed by your defense (or gained by opponent's offense, hence the "$Off$" prefix) on total yardage gained by your offense (right). Each non-scoring turnover forced by your defense mostly leads to higher offensive scoring, with majority of years exhibiting an approximately linear relationship. This is a reasonable finding due to each such turnover giving your offense the ball with a high potential for a good scoring opportunity. On the other hand, yards allowed per play by your defense exhibit a clear non-linear impact on your offense's ability to gain yards, which actually aligns well with the intuition. It is reasonable to expect for defenses allowing either extremely few ($<2.5$) or very many ($>5$) yards per play to take less of the game clock themselves, albeit for polar opposite reasons. Good defenses (low per-play yardage allowed) would quickly stop the opposing offense from scoring, and get the  ball over to their complementary offense (typically via a punt), putting the latter in a position to gain yards of their own. Bad defenses (high per-play yardage allowed) would let the opposing offense score points quickly, which still leads to their complementary offense getting the ball. Meanwhile, if your defense allows moderate per-play yardage ($2.5$ to $5$), that typically makes for longer offensive possessions by the opponent, taking  away from your own offensive unit the game time that could be used to gain more yards.

Note that, given the two-sided nature of game sports, the reasoning above could also be observed from the opposing team's viewpoint of their offensive side affecting their complementary defensive side. Each gain on offense could be considered as a loss on defense (e.g. a touchdown scored on offense is a touchdown allowed on defense) and vice versa (e.g. a sack gained by the defense is a sack allowed by the offense).

\subsubsection{Ranking shifts in points scored per game.}
\label{sec:RankingShifts}

To illustrate the mechanism behind adjustments for complementary football features, we showcase its impact on a team's offensive scoring ranking if one accounts for turnovers forced by that team's defense. Table \ref{tab:RankShiftsOffense} demonstrates the ranking shifts in points per game scored by offense in 2012 season when, on top of adjusting for strength of schedule and homefield factor, defense's ability to turn the ball over to the offense is accounted for. Oregon's scoring, while still remaining the best in FBS, dipped strongly (by $1$ point per game) once projected onto an average complementary unit, vastly due to their defense being Top-4 in forcing the non-scoring turnovers (same happened to the Washington offense, complemented by a Top-5 defense). On the other hand, Texas A\&M's offense climbed into Top-3 as it didn't benefit from as opportunistic of a defense (ranked outside Top-100). The biggest increases in projected points per game happened to Texas Tech and South Florida offenses, both working aside some of the worst defenses in terms of forcing turnovers (ranked 124th and 123rd, respectively, out of 125 teams). The intuition behind ranking shifts in other years was identical.

\begin{table}[h]
    \centering
    
 \begin{tabular}{rcc}
Team & Points Scored & Non-Scoring Turnovers Forced \\ & (Per Game) & (Per Possession) \\ \hline  
 \begin{tabular}{r} \\ Oregon \\ Louisiana Tech \\ Texas A\&M \\ Oklahoma State \\ Baylor \\ Clemson \\ Oklahoma \\ Alabama \\ West Virginia \\ Tennessee \\ Texas Tech \\ Georgia  \\ .... \\ South Florida \\ .... \\ Washington  \\ .... \\ \end{tabular} &
 \begin{tabular}{rrrr} Value & (Shift) & Rank & (Shift)\\ \hline 48.43 & ($\textcolor{red}{\Downarrow}$ 1.02) & 1 & ($\textcolor{blue}{0}$) \\ 47.12 & ($\textcolor{darkgreen}{\Uparrow}$ 0.19) & 2 & ($\textcolor{blue}{0}$) \\ 46.67 & ($\textcolor{darkgreen}{\Uparrow}$ 0.75) & 3 & ($\textcolor{darkgreen}{\Uparrow}$ 1) \\ 46.15 & ($\textcolor{darkgreen}{\Uparrow}$ 0.11) & 4 & ($\textcolor{red}{\Downarrow}$ 1) \\ 45.25 & ($\textcolor{darkgreen}{\Uparrow}$ 0.01) & 5 & ($\textcolor{blue}{0}$) \\ 41.02 & ($\textcolor{darkgreen}{\Uparrow}$ 0.08) & 6 & ($\textcolor{blue}{0}$) \\ 40.93 & ($\textcolor{darkgreen}{\Uparrow}$ 0.26) & 7 & ($\textcolor{darkgreen}{\Uparrow}$ 1) \\ 40.57 & ($\textcolor{red}{\Downarrow}$ 0.17) & 8 & ($\textcolor{red}{\Downarrow}$ 1) \\ 39.74 & ($\textcolor{red}{\Downarrow}$ 0.06) & 9 & ($\textcolor{blue}{0}$) \\ 39.69 & ($\textcolor{darkgreen}{\Uparrow}$ 0.58) & 10 & ($\textcolor{darkgreen}{\Uparrow}$ 2) \\ 39.44 & ($\textcolor{darkgreen}{\Uparrow}$ 1.22) & 11 & ($\textcolor{darkgreen}{\Uparrow}$ 4) \\ 38.98 & ($\textcolor{red}{\Downarrow}$ 0.19) & 12 & ($\textcolor{red}{\Downarrow}$ 1) \\ ... & ... \\21.99 & ($\textcolor{darkgreen}{\Uparrow}$ 1.16) & 87 & ($\textcolor{darkgreen}{\Uparrow}$ 6) \\ ... & ... \\23.72 & ($\textcolor{red}{\Downarrow}$ 1.02) & 80 & ($\textcolor{red}{\Downarrow}$ 5) \\ ... & ... \\ \end{tabular} & 
 \begin{tabular}{rr}  Value & Rank \\ \hline 0.19 & 4 \\ 0.13 & 52 \\ 0.08 & 114 \\ 0.12 & 75 \\ 0.12 & 61 \\ 0.12 & 67 \\ 0.09 & 111 \\ 0.17 & 13 \\ 0.10 & 97 \\ 0.09 & 108 \\ 0.06 & 124 \\ 0.16 & 23 \\ ... & ... \\0.06 & 123 \\ ... & ... \\0.19 & 5 \\ ... & ... \\ \end{tabular}
\end{tabular}
    \caption{\textbf{Final 2012 season rankings (and ranking shifts) in points scored per game by team's offense, adjusted for defense's ability to turn the ball over to team's offense.}
    }
    \label{tab:RankShiftsOffense}
\end{table}

\begin{table}[H]
    \centering
 \begin{tabular}{rcc} 
 Team & Points Allowed & Non-Scoring Turnovers Committed \\ & (Per Game) & (Per Possession) \\ \hline  
 \begin{tabular}{r} \\ Alabama \\ Notre Dame \\ Florida \\ Michigan State \\ Stanford \\ Brigham Young \\ South Carolina \\ Florida State \\ Rutgers \\ Kansas State \\ Louisiana State \\ Texas Christian  \\ .... \\ Pittsburgh \\ .... \\ Illinois  \\ .... \\ \end{tabular} &
 \begin{tabular}{rrrr} Value & (Shift) & Rank & (Shift)\\ \hline  8.70 & ($\textcolor{red}{\Uparrow}$ 0.42) & 1 & ($\textcolor{blue}{0}$) \\ 11.85 & ($\textcolor{red}{\Uparrow}$ 0.42) & 2 & ($\textcolor{darkgreen}{\Uparrow}$ 1) \\ 11.88 & ($\textcolor{red}{\Uparrow}$ 0.80) & 3 & ($\textcolor{red}{\Downarrow}$ 1) \\ 14.03 & ($\textcolor{red}{\Uparrow}$ 0.46) & 4 & ($\textcolor{darkgreen}{\Uparrow}$ 1) \\ 14.05 & ($\textcolor{red}{\Uparrow}$ 0.88) & 5 & ($\textcolor{red}{\Downarrow}$ 1) \\ 14.26 & ($\textcolor{red}{\Uparrow}$ 0.10) & 6 & ($\textcolor{blue}{0}$) \\ 15.74 & ($\textcolor{darkgreen}{\Downarrow}$ 0.60) & 7 & ($\textcolor{darkgreen}{\Uparrow}$ 1) \\ 15.97 & ($\textcolor{darkgreen}{\Downarrow}$ 0.44) & 8 & ($\textcolor{darkgreen}{\Uparrow}$ 1) \\ 16.56 & ($\textcolor{darkgreen}{\Downarrow}$ 0.31) & 9 & ($\textcolor{darkgreen}{\Uparrow}$ 2) \\ 16.84 & ($\textcolor{red}{\Uparrow}$ 0.88) & 10 & ($\textcolor{red}{\Downarrow}$ 3) \\ 17.06 & ($\textcolor{red}{\Uparrow}$ 0.25) & 11 & ($\textcolor{red}{\Downarrow}$ 1) \\ 17.31 & ($\textcolor{darkgreen}{\Downarrow}$ 0.60) & 12 & ($\textcolor{darkgreen}{\Uparrow}$ 2) \\ ... & ... \\25.07 & ($\textcolor{red}{\Uparrow}$ 1.26) & 51 & ($\textcolor{red}{\Downarrow}$ 8) \\ ... & ... \\27.41 & ($\textcolor{darkgreen}{\Downarrow}$ 1.27) & 64 & ($\textcolor{darkgreen}{\Uparrow}$ 5) \\ ... & ... \\ \end{tabular} & 
 \begin{tabular}{rr}  Value & Rank \\ \hline 0.09 & 21 \\ 0.10 & 32 \\ 0.09 & 19 \\ 0.09 & 14 \\ 0.10 & 29 \\ 0.13 & 70 \\ 0.13 & 77 \\ 0.15 & 92 \\ 0.13 & 69 \\ 0.08 & 9 \\ 0.10 & 33 \\ 0.16 & 104 \\ ... & ... \\0.06 & 2 \\ ... & ... \\0.20 & 124 \\ ... & ... \\ \end{tabular}
 \end{tabular}
    \caption{\textbf{Final 2012 season rankings (and ranking shifts) in points allowed per game by team's defense, adjusted for offense's ability to avoid turning the ball over to opposing offense.}
    }
    \label{tab:RankShiftsDefense}
\end{table}

We additionally provide Table \ref{tab:RankShiftsDefense} to demonstrate the  defensive ranking shifts for the same season. Symmetrically, the strongest positive (negative) impacts happened to the defensive units accompanied by offensive sides unable (able) to consistently take care of the ball and avoid turning it over to the opponent's offense. For example, teams like Illinois, Florida State, South Carolina, all sporting below-average offensive units in terms of ball security, had their defensive units move up the ranking. On the other hand, teams like Florida, Stanford, Pittsburgh, Kansas State, all experienced a considerable drop  defensive values due to being accompanied by offensive units that took care of the ball (all in the Top-30).

\subsubsection{Using adjusted statistics as features for game outcome prediction.}
\label{sec:PredictivePerformance}

In addition to exploration of the most impactful complementary football features, we compared predictive performances for statistics that were adjusted in various ways. Specifically, we pitched the following four methods against one another: no adjustment (classical averages), adjusting solely for strength of schedule and homefield factor (SOS+HF), adjusting also for efficiency-based complementary statistics (our main method), or for any complementary statistic.  The last approach is included for sanity check purposes, to demonstrate how reverse causality hurts predictive performance (see Section \ref{sec:ReverseCausality} for more details).

Table \ref{tab:PredictionEvaluationLinearMethods} shows the results of using logistic regression for binary game outcome prediction (whether team won or lost), and classic linear regression for modeling the score differential, respectively. As our five predictors we used offensive and defensive points-per-game margins for both teams, along with the homefield indicator. As prediction quality metrics, we used area under the curve (AUROC) \citep{hanley1982meaning} for binary game outcome, and mean absolute error (MAE) for score differential. To produce training errors, we utilized the full season data for each year in the 2009-2019 span. For test errors, we used the first 9 weeks to train the model and subsequently yield predictions for the rest of the season (using other cut-offs, e.g. 8, 10, 11 weeks, led to similar results).

\begin{table}[]
    \centering
    \resizebox{\textwidth}{!}{
 \begin{tabular}{rcc}
 & Binary Game Outcome & Score Differential \\
Adjustment &  (Won/Lost, AUROC)  & (MAE)  \\
\hline
\begin{tabular}{r}
\\
None \\
SOS+HF \\
SOS+HF \& Efficiency Cmpl. \\
SOS+HF \& Any Cmpl. \\
\end{tabular} &

\begin{tabular}{rr} 
Training & Test \\
\hline
0.790 (0.017) & 0.704 (0.028) \\
0.810 (0.017) & 0.715 (0.026) \\
 0.808 (0.016) & 0.717 (0.033) \\
 0.790 (0.019) & 0.707 (0.026) \\
\end{tabular} &

\begin{tabular}{rr} 
Training & Test \\
\hline
14.76 (0.57) & 17.39 (1.20) \\
13.78 (0.58) & 17.12 (1.18) \\
13.83 (0.57) & 17.13 (1.18) \\
14.58 (0.58) & 17.53 (0.93) \\
\end{tabular}

\end{tabular}}
 \caption{\textbf{Predicting binary game outcomes (won/lost) via logistic regression, and numerical scoring differentials via linear regression, across 2009-2019 seasons by using variously adjusted points per game margin stats.} Training metric obtained from fitting the model to entire season, while test metric is obtained from using only first 9 weeks to predict the rest of the season.}
    
    \label{tab:PredictionEvaluationLinearMethods}
\end{table}

Performances are mostly similar across the methods, with all the average metrics being within less than two standard deviations of one another. Nonetheless, SOS+HF and our method consistently show better average performances (higher AUROC, lower MAE values). When adjusting for any complementary football features, rather than only efficiency-based ones, the performance was as suboptimal as the classical averages, confirming the dangers of reverse causality in case of plainly using the game totals.

%% file: Arxiv_Submission/5_Discussion.tex
\section{Discussion and future work}
\label{sec:Discussion}

We have applied several group penalization techniques in combination with natural cubic splines 
to detect the most consistent features of American college football in terms of the impact on the complementary side of your team (e.g. your offense affecting your defense, or vice versa). Partially constrained optimization was implemented to also guarantee the adjustment for strength of the opponent and homefield factor. Issue of reverse causality for internal dynamics of a football game was discussed, along with several remedies, the most feasible being to focus solely on the efficiency-based statistics (e.g. yards per play, points per possession, etc). Benefits of using efficiency statistics were shown in evaluating game outcome predictions based on various adjustments, where efficiency-based complementary statistics outperformed arbitrary complementary statistics, with the latter suffering from reverse causality more heavily.

Among the findings, we unsurprisingly showed the non-scoring turnovers (possession percentage) forced by your defensive unit as the most consistent feature in positively impacting the scoring by your offensive unit. This is reasonable due to such turnovers giving the your offensive side an extra possession along with the potential for a good starting field position. Hence, the ranking shifts in points or touchdowns scored per game that resulted from this adjustment tended to penalize offensive units complemented by a defensive side that forced many non-scoring turnovers (conversely, in points or touchdowns allowed, most penalized were defensive units complemented by offensive sides that committed few turnovers).

On the other hand, team's defensive statistic mostly affecting the total yardage gained by the offensive unit was the yards allowed per play. This effect was non-linear, with defenses allowing a moderate per-play yardage (about 3 to 5) being less conducive to their complementary offensive units gaining yards in that game, while defenses on more extreme ends of the spectrum - allowing either tons (e.g. $\geq 6$) or very few (e.g. $\leq 2.5$) yards per play - tend to give more opportunities for their offense to gain yards. As thoroughly discussed in Section \ref{sec:SelectedEfficiencyStatistics}, this finding is intuitive due to the moderate yards per play generally indicating the inability of your defense to stop the opposing offense from controlling the game clock, hence preventing your offense from gaining yards by keeping it on the sidelines.

In future work, we plan on gaining access to the play-by-play data and incorporating it into the analysis for the purpose of alleviating the reverse causality issue, resulting into more accurate evaluation of complementary football feature effects. Although using solely efficiency-based statistics leads to reasonable results, it constitutes only a partial solution for scenarios of limited data access (e.g. game-by-game total statistics rather than play-by-play) and could still suffer from endogeneity, which manifests in lack of improvement over pure strength-of-schedule and homefield factor (SOS+HF) adjustment for game outcome predictions. Lastly, 
one may consider trying to find strong exogenous instrumental variables, which has been a notoriously difficult task.

%% file: Arxiv_Submission/6_Acknowledgments.tex
\section*{Acknowledgments}
\label{sec:Acknowledgments}

 The author is grateful to \url{www.sports-reference.com/cfb/} and \url{www.cfbstats.com/} for providing open access comprehensive game-by-game college football data.

%% file: Arxiv_Submission/7_Declaration_of_Interest.tex
\section*{Declaration of interest}
\label{sec:DeclarationOfInterest}

We confirm that there are no known conflicts of interest associated with this publication and
there has been no financial support that could have influenced its outcome.

%% file: Arxiv_Submission/8_Appendices.tex
\section*{Appendix}

\subsection{Feature sets.}
\label{sec:FeatureSets}

\subsubsection{All features (not only efficiency-based).}
\label{sec:AllFeatures}

After disposing of strong pairwise correlations (e.g.  $Off.Pts$, $Off.TD$, $Off\_w\_ST.TD$ all have a correlation of $0.95+$ with the more all-encompassing and ubiquitously used $Off\_w\_ST.Pts$, etc), perfect multi-collinearities (e.g. instead of including passing attempts, completions and incompletions, only retaining the individual breakdown stats of completions and incompletions), unneeded statistic breakdowns (e.g. there is no difference to the complementary side if it was an individual or assisted tackle, whether the turnover was an interception or a fumble, etc), and all low-denominator efficiency statistics (field goal, red zone, fourth down conversion percentages, punt and return yards per attempt, etc), we were left with the features in Table \ref{tab:SuppFeatureSetTab1}.  Note that the overall turnovers by offense and offensive special teams are not included because they cause multi-collinearity with non-scoring turnovers and defensive (plus defensive special teams) touchdowns, with that breakdown warranted due to difference of the impact on the complementary side.

\begin{table}[]
    \centering
    \resizebox{\linewidth}{!}{%
    \begin{tabular}{rr}
    Name & Description \\
    \hline
       $Off.Tot$ &  Total first downs \\
       $Off.Cmp$ & Pass completions  \\
     $Off.Incmp$   & Pass incompletions \\
       $Off.Pct$   & Pass completion percentage \\
      $Off.Pass.Yds$  & Total passing yards \\
      $Off.Pass.Avg$   & Average yards per pass attempt \\
      $Off.Rush.Att$    & Rush attempts \\
       $Off.Rush.Yds$   & Total rushing yards \\
       $Off.Rush.Avg$   & Average yards per rush attempt \\
       $Off.FGM$        &   Field goals made \\
       $Off.FGF$     &   Field goals failed \\           
     $Off.X3rdDown..Conversions$ & Third downs converted \\
     $Off.X3rdDown..Failed$   &   Third downs failed \\
     $Off.X3rdDown..Conversion..$ & Third down conversion percentage \\
     $Off.X4thDown..Conversions$ & Fourth downs converted \\
     $Off.X4thDown..Failed$  & Fourth downs failed \\
     $Off.RedZone..TD$   & Red zone touchdowns \\
     $Off.RedZone..FG$   & Red zone field goals \\
     $Off.XPF$     & Extra points failed \\
     $Off.ST.Punting.Punts$  & Punts \\
     $Off.ST.Return.TD$    & Return touchdown (punt or kickoff) \\
     $Off.ST.Return.Yds$   & Total return yards (punt or kickoff) \\
     $Off.ST.Return.Ret$   & Total returns attempted (punt or kickoff) \\
     $Def.Sfty$         & Safeties \\
     $Def.QB.Hurries$   &   Quarterback hurries \\   
     $Def.KicksPunts.Blocked$   & Kicks or punts blocked \\
     $Def.Tackles.Tot$   & Total tackles \\
     $Def.Tackles.Loss$    & Tackles for loss \\
     $Def.Tackles.Sk$    & Sacks \\
     $Def.Int.PD$       & Passes defended \\
     $Def.Fumbles.FF$   & Fumbles forced \\
     $Off\_w\_ST.Pts$     & Points scored\\
     $Off\_w\_ST.Tot.TO.NonScoring$ & Non-scoring turnovers \\
     $Def\_w\_ST.TD$   & Touchdowns scored (by defensive side ) \\
     $Def\_w\_ST.Yds$  & Yards gained (by defensive side) \\
  
    \end{tabular}
    }
    \caption{\textbf{Glossary of all original complementary football features.} $Off$ stands of "offense-only", $Def$ - "defense-only", $Off.ST$ - "offensive special teams", $Off\_w\_ST$ - "offense with offensive special teams", $Def\_w\_ST$ - "defense with defensive special teams".}
    \label{tab:SuppFeatureSetTab1}
\end{table}

\subsubsection{Efficiency-based features.}
\label{sec:OnlyEfficiencyBasedFeatures}

Just like in case of the original feature set, we dispose of strong pairwise correlations, perfect multi-collinearities, unneeded statistic breakdowns, and all low-denominator efficiency statistics. As described in the main manuscript, for each statistical category we had to decide on whether to use a per-play or per-possession efficiency. We've selected to look at plays that terminate a possession, e.g. turnover, points scored, punt, on per-possession basis, making them into possession percentages. For example, $Off\_w\_ST.NonScoring.TO.PossPct$ corresponds to the percentage of possessions that ended in the offense turning the ball over to their defense (as opposed to turning it over for an instant score by the opposing defense, which is captured via $Def\_ST.TD.PossPct$). On the other hand, virtually all other metrics were treated on per-play basis, including yards or first downs gained, overall tackles, tackles for loss, forced fumbles. Lastly, one can notice such statistical categories as "sacks to pass attempts ratio", which were defined in an effort to adjust for effects of pass-heavy play calling being conducive to such defensive plays as sacks or quarterback hurries. While it's understood that sack (or quarterback hurry) is not recorded as a passing attempt, and ideally one would need the number of quarterback dropbacks to make it a proper efficiency percentage, but due to data limitations this was a reasonable alternative. In similar vein, percentage of forced fumbles by defense was only counted for rushing attempts and pass completions, meaning the plays where fumble was possible. Table \ref{tab:SuppFeatureSetTab2} demonstrates all the efficiency-based features considered.

\begin{table}[]
    \centering
    
    \resizebox{\linewidth}{!}{%
    \begin{tabular}{rr}
        Name & Description \\
        \hline
 $Off.Pct$  & Pass completion percentage \\
 $Off.Pass.Avg$   & Average yards per pass attempt \\
 $Off.Rush.Avg$   & Average yards per rush attempt \\
 $Off.Tot.Avg$ & Average yards per play \\
 $Off.X1stDowns.Avg$ & Average first downs gained per play \\               
 $Off.X3rdDown..Conversion..$ &  Third down conversion percentage \\
 $Def.KicksPunts.Blocked.Pct$ &  Percentage of kicks and punts blocked \\
 $Def.Tackles.Tot.Avg$ &   Tackles per play \\
 $Def.Tackles.Loss.Avg$ &  Tackles for loss per play \\           
 $Def.QB.Hurries.To.Pass.Att.Ratio$  & Ratio of QB hurries to pass attempts \\
 $Def.Tackles.Sk.To.Pass.Att.Ratio$  & Ratio of sacks to pass attempts \\
 $Def.Int.PD.Avg$  & Defended passes per pass attempt \\
 $Def.Fumbles.FF.Avg$  & Forced fumbles per play \\
 $Off\_w\_ST.Punt.Sfty.PossPct$     & Punt or safety \\
 $Off\_w\_ST.Tot.TO.NonScoring.PossPct$ & Non-scoring turnovers  \\
 $Off\_w\_ST.Pts.PerPos$  & Points scored   \\
 $Def\_w\_ST.TD.PossPct$ & Touchdowns scored (by defensive side)  \\
    \end{tabular}
    }
    \caption{\textbf{Glossary of all efficiency-based complementary football features.} $Off$ stands of "offense-only", $Def$ - "defense-only", $Off.ST$ - "offensive special teams", $Off\_w\_ST$ - "offense with offensive special teams", $Def\_w\_ST$ - "defense with defensive special teams", $PossPct$ - "possession percentage", $Avg$ - per-play average, $PerPos$ - "per possession".}
    \label{tab:SuppFeatureSetTab2}
\end{table}